\documentstyle[aps,twocolumn,prb,epsf]{revtex}

\begin{document}
\draft

\twocolumn[\hsize\textwidth\columnwidth\hsize\csname @twocolumnfalse\endcsname

\title{Finite-temperature properties of the two-orbital Anderson model}

\author{Luis Craco}

\address{Instituto de F\'{\i}sica Gleb Wataghin - UNICAMP, C.P. 6165, 
13083-970 Campinas - SP, Brazil}

\date{\today}

\maketitle

\widetext

\begin{abstract}
\noindent
The metallic phase of the two-orbital Anderson lattice 
is study in the limit of infinite spatial dimensions, where a 
second order perturbation treatment is used to 
solve the single-site problem.  Using this approximation, in the  
Kondo regime, we find that the finite temperature properties of  
the conduction electrons exhibit  the same behaviour 
as observed in the metallic phase of the two-channel Kondo lattice.
Possible connections between these two models are discussed.  
\end{abstract}

]    

\narrowtext

The normal metallic state of a number of three-dimensional heavy-fermion 
compounds can be quite well described by the Landau Fermi  liquid theory.
At zero temperature $(T)$ the heavy-fermion quasi-particles have an infinite
lifetime and a large effective mass. As a  consequence of this {\it coherent} 
regime the heavy-fermion compounds  present at low $T$  a very large 
electronic specific heat coefficient $\gamma (T) = C(T)/T$.~\cite{Fulde-book}  
The physical properties of these materials are related to strongly correlated
electrons in $4f/5f$ orbitals, and a proper model Hamiltonian for the
description of these properties is the Anderson Hamiltonian.~\cite{Fulde-book}

The Fermi liquid regime in the single impurity 
Anderson model can be identified by the behaviour of the $f$-electron
self-energy $\Sigma^f (\omega)$ at low temperatures near of the Fermi 
level $(\varepsilon_F)$. The $f$-electron self-energy in the single  
impurity Anderson model is ${\bf k}$-independent and respects the 
Fermi liquid requirements.~\cite{CG-H} 
Performing perturbation expansion in the $f$-electron
Coulomb interaction $(U)$ at low $T$ and near $\varepsilon_F$
Yamada showed that ${\rm Re} \Sigma^f (\omega) \approx -\omega$ and 
${\rm Im} \Sigma^f (\omega) \approx - (\omega^2 + (\pi T)^2)$.~\cite{Yamada}
For the lattice case and  
in the infinite dimensional limit, where $\Sigma^f (\omega)$ is also 
${\bf k}$-independent,~\cite{MetVol,Georges} it was shown by Georges 
{\it et al.}~\cite{GKS} that the metallic phase of the 
Periodic (one-orbital) Anderson model can be  
a Fermi liquid. The behaviour of the real and imaginary 
part of $\Sigma^f (\omega)$ in $d=\infty$ was studied by Schweitzer and
Czycholl~\cite{SC} by a  self-consistent second order perturbation  in $U$.
In this case the imaginary part of $\Sigma^f (\omega)$
vanishes near $\varepsilon_F$ in accordance with the Luttinger theorem, 
and the real part has a negative slope in the same region. 

The investigations  presented above were done only for one-channel
versions of the Anderson Hamiltonian. The multichannel one impurity 
Anderson model, where a localised $f$-electron hybridises  with several 
conduction bands (orbitals), was studied in the large U 
limit.~\cite{crkw,Kroha} In this limit the model shows non-Fermi 
liquid behaviour. A non-Fermi liquid behaviour is also present  
in the multichannel Kondo impurity problem. This problem  was 
introduced by Nozi\`eres and Blandin,~\cite{Nozieres} and the 
exact solution was obtained by Andrei and Destri~\cite{Andrei} 
and Wiegmann and Tsvelick~\cite{Wiegmann} in terms of the Bethe {\it Ansatz}.
The non-Fermi liquid behaviour in the multichannel Kondo impurity
model comes from the overcompensation of the Kondo
spins by the conduction electrons.~\cite{Schlottmann}  The overcompensation 
mechanism was also used to explain the non-Fermi liquid behaviour in the 
two-channel Kondo (tCK) lattice ~\cite{Jarrell96,Jarrell97} and the two-channel
Anderson (tCA) lattice~\cite{Anders} in the Kondo limit. 

The conclusion about the non-Fermi liquid regime in the tCK lattice 
was made by Jarrell {\it et al.}~\cite{Jarrell96} looking at the one-particle 
properties of the conduction electrons.  In this case,  the real part 
of the self-energy of the conduction electrons $\Sigma^c (\omega)$ 
present a positive slope near the Fermi energy. The imaginary part of 
$\Sigma^c (\omega)$ instead going to zero in a quadratic way as in 
Fermi liquid systems it goes away from zero as 
\mbox{$\omega, T \rightarrow 0$}. Despite of these properties the system is 
metallic and the single-particle density of states (DOS) 
of the conduction electrons has a finite value at the Fermi level.
This metallic regime ({\it incoherent metal}) was used  
to explain the physical properties of a number of heavy 
Fermion compounds where the Fermi liquid paradigm can not be 
applied.~\cite{Jarrell96,Cox}
For example, the incoherent metal regime have been used to explain  
the unusual resistivity of $UBe_{13}$.~\cite{Anders} 

In addition, in Ref.~[14] the $t$CK model was studied by means 
of quantum Monte Carlo (QMC) simulation in the limit of high dimensions. 
It is well known that QMC method provides very accurate results at 
intermediate and high temperatures. However, it does not provide any 
information about the explicit form of the self-energy of the problem. 
For example, in Jarrell's {\it et al.}~\cite{Jarrell96} work the 
single particle self-energy was obtained by inverting the relation 
$G^c(\omega)=D (\omega -\Sigma^c(\omega))$~\cite{D} for different 
temperatures. It must be noted that the nonanalytic form of 
$\Sigma^c(\omega)$ was postulated from these numerical results. Thus, in order 
to understand the origin of the incoherent metal regime in $t$CK lattice, 
it is important to find out a model which describes these novel properties 
and allow us to obtain an explicit form for $\Sigma^c(\omega)$. Here, we 
address precisely this problem: the study of a multi-orbital model which 
correctly takes into account the incoherent behaviour of the $t$CK 
lattice and provides an explicit form for the self-energy of the 
conduction electrons of each orbital.

In this letter we study a multi-orbital Anderson lattice ($m$OA) model
in the high dimension $(d=\infty)$ limit. We discuss the formal exact 
solution of this multi-orbital Hamiltonian as well as the finite temperature 
properties of the two-orbital version of this problem. In this case a 
second order perturbation treatment is applied to solve the impurity 
problem in the presence of $U$. As we have mention before, we find that 
the conduction electrons show the same behaviour as those obtained by 
Jarrell {\it et al.}~\cite{Jarrell96} for the tCK lattice.

The $m$OA model consists of the usual $f$-electron Hamiltonian of the 
periodic Anderson lattice model, $m$ identical orbitals of noninteracting  
$c$-electrons and the local hybridisation between $f$ and $c$ electrons.  
The complete Hamiltonian can be written as 

\begin{eqnarray}
\label{eq:model}
H & = & -\frac{t^\ast}{2\sqrt d} \sum_{\langle ij\rangle\sigma\alpha} 
c_{i\alpha\sigma}^\dagger c_{j\alpha\sigma}   
+ E\sum_{i\sigma} n_{i\sigma}^f + 
U\sum_i n^{f}_{i\uparrow} n^{f}_{i\downarrow} \nonumber \\ && +
\sum_{i\sigma\alpha} V_{\alpha \sigma} \left( c_{i\alpha\sigma}^\dagger 
f_{i\sigma} + f_{i\sigma}^\dagger c_{i\alpha\sigma} \right) \;,
\end{eqnarray}
where $c_{i\alpha\sigma}^\dagger$ $(c_{i\alpha\sigma})$  creates (destroys)
a conduction electron on site $i$ and orbital $\alpha=1,2,...,m$ of spin 
$\sigma$, and $f_{i\sigma}^\dagger$ $ (f_{i\sigma})$, creates (destroys) a 
localised $f$-electron on site $i$ of spin $\sigma$. The sites $i$ form an 
infinite dimensional hyper-cubic lattice and the hopping is limited to the 
nearest neighbours. The scaled hopping integral $t^\ast = 1$ determines the 
energy unit. The hybridisation term $V_{\alpha\sigma}$is site independent, 
but it can have different values for different orbitals or spin directions. 

To obtain the  formal exact solution of the $m$OA model it is 
convenient to define a new set of conduction electron operators:
$\{a_{i 1 \sigma}^\dagger, a_{i 2 \sigma}^\dagger ... a_{i m 
\sigma}^\dagger\}$. In this case 
\mbox{$a_{i 1 \sigma}^\dagger \equiv 1/{\bar V_\sigma} 
\sum_\alpha V_{\alpha\sigma} c_{i\alpha \sigma}^\dagger$} and the 
remainder $a_{i \alpha \sigma}^\dagger$ operators are written in order to 
preserve the definition of $a_{i1\sigma}^\dagger$ and the fermion 
commutation relations. The normalisation factor is  
$\bar V_{\sigma} =\sqrt{ \sum_\alpha V_{\alpha\sigma}^2}$. 

In the new representation the $m$OA Hamiltonian 
(Eq.~(\ref{eq:model})) is written as  

\begin{eqnarray}
\label{eq:mod-a}
H & = & -\frac{t^\ast}{2\sqrt d} \sum_{\langle ij\rangle\sigma\alpha} 
a_{i\alpha\sigma}^\dagger a_{j\alpha\sigma}   
+ E\sum_{i\sigma} n_{i\sigma}^f + 
U\sum_i n^{f}_{i\uparrow} n^{f}_{i\downarrow} \nonumber \\ && +
\sum_{i\sigma} \bar V_{\sigma} \left( a_{i1\sigma}^\dagger f_{i\sigma}
+ f_{i\sigma}^\dagger a_{i1\sigma} \right) \;.
\end{eqnarray}

It is clear from  Eq.~(\ref{eq:mod-a}) that the multi-orbital problem 
for the $f$-electrons is  reduced to a one-orbital problem with a 
renormalised hybridisation $\bar V_{\sigma}$. As the local 
approximation for the Anderson  (one-orbital)
lattice is exact in the limit of infinite 
dimensions,~\cite{Georges,Consiglio} the formal exact solution for the  
$f$-electron one-particle Green's function in $d\rightarrow \infty$ 
is given by~\cite{Georges}

\begin{equation}
\label{eq:gfk1}
G_{{\bf k} \sigma}^f (i\omega_n)^{-1} = 
i\omega_n - \Sigma_{\sigma}^f (i\omega_n) -   
\frac{\bar V_\sigma^2}{i\omega_n  - \epsilon_{\bf k} } \;.
\end{equation}

As we intend in this work to clarify the physical origin of the 
incoherent properties of the tCK model,
the conduction electron Green's function of each 
particular orbital must be known. In the context of the mapped 
Hamiltonian (Eq.~(\ref{eq:mod-a})) the $\alpha$-orbital Green's function
$G_{ij\sigma}^{\alpha c}(\tau) \equiv -\langle \hat T 
c_{i\alpha\sigma}^{}(\tau) c_{j\alpha\sigma}^{\dagger}(0) \rangle$ is 
obtained when the $c_{i\alpha\sigma}$ operators are written in terms of the  
$a_{i\alpha\sigma}$ operators, and it is straightforward to show that

\begin{equation}
\label{eq:gcii}
G_{ii\sigma}^{\alpha c} (i\omega_n) =
\left(1- \frac{V_{\alpha \sigma}^2}{\bar V_\sigma^2} \right) D(i\omega_n) +
\frac{V_{\alpha \sigma}^2}{\bar V_\sigma^2} 
D(i\omega_n - \bar V_\sigma^2 {\cal G}_{\sigma}) \;,  
\end{equation}
where ${\cal G}_{\sigma} (i\omega_n)^{-1} \equiv i\omega_n - 
\Sigma_{\sigma}^f (i\omega_n)$ and $D(z)$ is the Hilbert transform of the 
uncorrelated density of states of the conduction 
band $\rho_0(\epsilon)$.~\cite{D} For a hypercubic lattice in $d=\infty$  
$\rho_0(\epsilon) = (1/\sqrt \pi) \exp (-\epsilon^2)$.

The two terms of Eq.~(\ref{eq:gcii}) can be easily understood from  
the transformation discussed above.   
The first term of the right-hand side of this equation describes  
the contributions of the $(m-1)$ free orbitals and the second term
is related to the one-orbital Anderson problem of Eq.~(\ref{eq:mod-a}).
It is important to notice that the formal exact solution of the $m$OA model
(Eqs.~(\ref{eq:gfk1}) and~(\ref{eq:gcii})) is completely general concerning
the number of orbitals as well as the values of the hybridisation on different
orbitals. 

In principle, it is possible to assume that the {\it scattering processes} 
in each orbital can be described by means of a 
self-energy.~\cite{Jarrell96,Consiglio} By definition, the 
$\alpha$-orbital  Green's function can be written in terms of the 
self-energy of the conduction electrons  
$\Sigma_{{\bf k} \sigma}^{\alpha c} (i\omega_n)$ as
     
\begin{equation}
\label{eq:gck1}
G_{{\bf k} \sigma}^{\alpha c} (i\omega_n)^{-1} \equiv 
i\omega_n  - \epsilon_{\bf k} - 
\Sigma_{{\bf k} \sigma}^{\alpha c} (i\omega_n)\;.
 \end{equation} 
In order to recover Eq.~(\ref{eq:gcii}) by summing Eq.~(\ref{eq:gck1}) 
over momentum, the self-energy of the $\alpha$-orbital conduction 
electrons must be given by

\begin{eqnarray}
\label{eq:selfc}
\Sigma_{{\bf k} \sigma}^{\alpha c} (i\omega_n) & = &  
V_{\alpha \sigma}^2 {\cal G}_\sigma (i\omega_n) 
\nonumber \\ & \times & \left\{ 1 +
\frac{ (\bar V_\sigma^2 - V_{\alpha\sigma}^2) {\cal G}_\sigma (i\omega_n)}
{i\omega_n  -  (\bar V_\sigma^2 - V_{\alpha\sigma}^2) 
{\cal G}_\sigma (i\omega_n) - \epsilon_{\bf k} } \right\} \;.
\end{eqnarray}

The requirement that in the $d \rightarrow \infty$ limit the local 
interactions give rise to a ${\bf k}$-independent self-energy for the 
$f$-electrons is satisfied for the $m$OA lattice 
(see Eq.~(\ref{eq:gfk1})). However, a ${\bf k}$-dependent self-energy 
for the $\alpha$-orbital conduction electrons is obtained.
 
The first term of $\Sigma_{{\bf k} \sigma}^{\alpha c} (i\omega_n)$, 
$V_{\alpha \sigma}^2 {\cal G}_\sigma (i\omega_n)$, is the self-energy of
one-orbital Anderson lattice. This term is completely local and describes 
the scattering 
processes in the $\alpha$-orbital. The second term is relevant only for 
multi-orbital systems, where $\bar V_\sigma^2 \ne V_{\alpha\sigma}^2$. 
As $(\bar V_\sigma^2 - V_{\alpha\sigma}^2)$ is related to 
all other orbitals different from  $\alpha$, the ${\bf k}$-dependent 
term can be considered as an {\it effective correction}
to the one-orbital self-energy. This effective correction describes the 
scattering process that happens in the remaining orbitals. 
 
Different methods can be used to solve the single site problem 
in the $d=\infty$ limit.~\cite{Georges,Craco2} One of these methods is    
the iterative perturbation theory (IPT).~\cite{IPT} IPT    
shows to be a good approximation for describing the Fermi liquid 
properties of the Hubbard model.~\cite{zrk}  For the particular 
case of the periodic Anderson model the IPT results are in good agreement 
with exact diagonalisation~\cite{Georges} and QMC.~\cite{Jarrell93} 
The IPT method is normally applied for values of $U/t^\ast < 3$, 
however it  has the property of correctly taking into account 
the limit of $V \rightarrow  0$.~\cite{Georges} This allow us to 
study the one-particle properties of the $m$OA lattice near the 
Kondo limit. 

\begin{figure}
\epsfxsize=3.5in
\epsffile{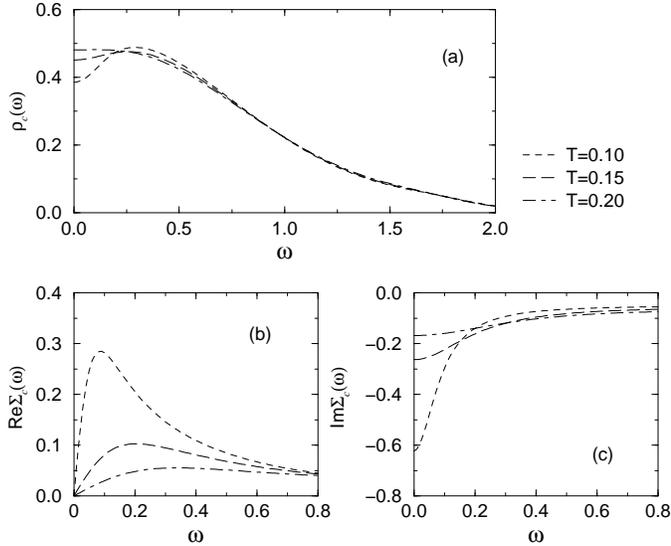}
\caption{One-particle properties of the $c$-electrons
for $m=2$, $V=0.3535$, $U=2.0$ and three different temperatures. 
(a) Density of states, (b) Real and (c)
Imaginary parts of the self-energy, respectively.}
\label{fig:2}
\end{figure}

Let us now turn our attention on the spectral properties of the conduction 
electrons of the two-orbital Anderson (tOA) model.  The one-particle 
properties of $f$-electrons will not be considered here because they are 
well known from studies of the periodic Anderson 
model.~\cite{Georges,SC,Jarrell93} In our study we have chosen 
$U=2.0$ and $V=0.3535$. Note that  $V=0.3535$
for $m=2$ means $\bar V=0.5$.

In Fig.~\ref{fig:2}-a we display the single particle density of states
(DOS) for the conduction electrons of tOA model. The DOS 
has a finite value at the Fermi level for all temperatures.
Such metallic behaviour is related to charge fluctuations
and to the contribution of the non-hybridised electrons.
The {\bf k}-independent real and imaginary part of the $c$-electrons 
self-energy $\Sigma^c (\omega) \equiv V^2 {\cal G} (\omega)$
(see Eq.~(\ref{eq:selfc})) is plotted in  Fig.~\ref{fig:2}-b
and~\ref{fig:2}-c, respectively. As one can see in  Fig.~\ref{fig:2}-b,
the real part of the self-energy exhibits a positive slope near the Fermi 
level and this slope decreases with increasing temperature. It is well known
that for a Fermi liquid system the slope of $Re \Sigma^c (\omega)$ must be
negative. Therefore, as it was pointed out in Ref.~[14]   
 the positive slope observed in Fig.~\ref{fig:2}-b describes the 
breakdown of the quasiparticle concept. Non-Fermi liquid properties 
can be also observed in the imaginary part of the self-energy (see 
Fig.~\ref{fig:2}-c). From this figure it is clear that 
$Im \Sigma^c (\omega)$ does not approach to the Fermi liquid form
$Im \Sigma (\omega) \approx -(T^2 + \omega^2)$ as 
$\omega,T \rightarrow 0$. Note that similar behaviour as presented 
in Fig.~\ref{fig:2} has been observed in the one-particle
properties of the conduction electrons of tCK lattice.~\cite{Jarrell96}  

It is worth noticing that the incoherent properties shown in 
Fig.~\ref{fig:2}-b and~\ref{fig:2}-c are closely related to the definition 
of $\Sigma^c(\omega)$. Concerning this point, we follow the idea of Jarrell 
{\it et al.}~\cite{Jarrell96} where the self-energy is defined in order to 
account for the Hilbert transform of the Green's funtion 
of the conduction electrons in each channel.
It is well known that $\Sigma^f(\omega)$ (see Eq.~(\ref{eq:gfk1}))
shows Fermi liquid behaviour at low temperatures near to the Fermi 
level.~\cite{SC} From Eq.~(\ref{eq:selfc}) one can easily conclude that 
the Fermi liquid properties of the $f$-electrons imply incoherent
properties for the conduction electrons of the periodic 
Anderson model as well as the $m$OA model. Hence, to decide whether 
a model of two or more different particles is in a coherent regime or 
not, it is important to consider the behaviour of all particles. 

\begin{figure}
\epsfxsize=3.in
\epsffile{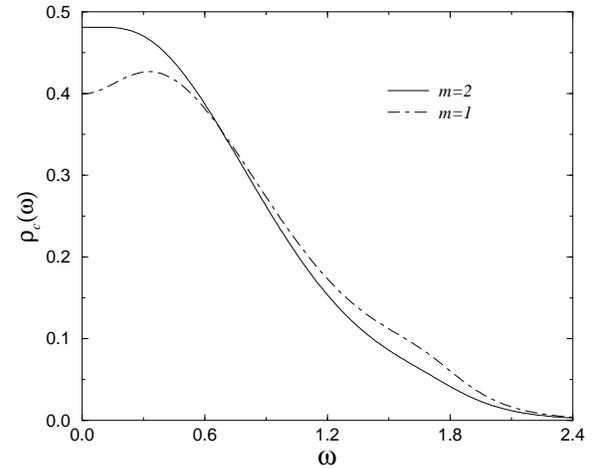}
\caption{DOS of $c$-electrons at $T=0.20$ for
$m=2$ (solid line) and $m=1$ (dot-dashed line).} 
\label{fig:3}
\end{figure} 

In Fig.~\ref{fig:3} we compare the DOS for the conduction electrons
of the tOA model with those of the periodic Anderson model. In this 
figure we consider two different values of hybridisation:  
$V=0.3535$ $(m=2)$ and  $V=0.5$ $(m=1)$, such that the 
correlation effects are taken into account at the same level for both 
systems. This allow us to study the effect of the non-hybridised 
particles in the two-orbital system. For $T=0.20$ the contribution 
of this particles completely suppresses the hybridisation gap and a flat 
region in the DOS of $m=2$ is observed near the Fermi level. Despite the 
differences in the Hamiltonian parameters and the method used 
to solve the impurity problem similar flat behaviour
have  been reported in Ref.~[14] for tCK model. 

Let us now proceed to understand the similarities between our
results and those of tCK model.~\cite{Jarrell96} 
In the Kondo limit the $m$OA model (Eq.~(\ref{eq:model}))  
is mapped onto a multi-orbital Kondo ($m$OK) model. 
In order to show this result one can apply the resolvent 
perturbation theory~\cite{Keiter} in the subspace defined by the 
$a_{i1\sigma}$ and $f$ operators, see Eq.~(\ref{eq:mod-a}). If the 
hybridisation is the same for both spin directions the $m$OK 
Hamiltonian is written as   

\begin{eqnarray}
\label{eq:Kmod}
H_{Kondo} 
& = & -\frac{t^\ast}{2\sqrt d} \sum_{\langle ij\rangle\sigma\alpha} 
c_{i\alpha\sigma}^\dagger c_{j\alpha\sigma} + 
k \sum_{i\alpha} V_{\alpha}^2 {\bf S}_i \cdot {\bf s}_{i\alpha} 
 \nonumber \\ & + & 
k \sum_{i\alpha \ne \alpha'} V_{\alpha}V_{\alpha'} [ S_i^z
(c_{i\alpha\uparrow}^\dagger c_{i\alpha'\uparrow} -
c_{i\alpha\downarrow}^\dagger c_{i\alpha'\downarrow})  \nonumber \\ & + &  
S_i^+ c_{i\alpha\downarrow}^\dagger c_{i\alpha'\uparrow} + 
S_i^- c_{i\alpha\uparrow}^\dagger c_{i\alpha'\downarrow} ] \;,
\end{eqnarray}
where $k \equiv U/(|E|(|E|+U))$~\cite{Fulde-book} and 
${\bf S}_i$, ${\bf s}_{i\alpha}$ are the Kondo and the 
$\alpha$-orbital conduction electron spin operators, respectively.
Note that Eq.~(\ref{eq:Kmod}) is precisely the Schrieffer-Wolf 
transformation of Eq.~(\ref{eq:model}).

The first two terms on the right hand side of Eq.~(\ref{eq:Kmod})
can be considered as the simplest generalisation of the {\it one} and 
{\it two-channel} Kondo models into a multi-channel one.
The remaining terms are related to non-diagonal exchange 
processes between the impurity and the conduction electrons in 
the different orbitals. The relevant contributions of these 
non-diagonal processes will appear only in fourth order in 
the hybridisation term, see Eq.~(\ref{eq:selfc}). It is clear that  
the fourth order contributions act only as small corrections in the 
Kondo limit, $(kV_\alpha V_{\alpha'})^2 \ll 1$. This explain the 
agreement between our results and those of tCK model. 

Summarising, the simplest extension of the periodic Anderson model 
into a multi-orbital Anderson ($m$OA) model is introduced 
for the first time in this letter.
The model is studied in the limit of high dimensions, where a second 
order perturbation treatment (IPT) is used to solve the impurity 
problem. Using this approximation at finite temperatures we find
that the single-particle properties of the conduction electrons
in the Kondo regime for the two-orbital Anderson model 
are the same as those for the two-channel Kondo lattice.  
We have explained this agreement in terms of the irrelevance of the 
non-diagonal exchange processes in the Kondo limit.
Finally, we wish to point out 
the importance of our results since they open a new possibility of 
study the $m$CK problem in the limit of high dimensions. 
     
\acknowledgments
The author wishes to acknowledge, respectively,  M. A. Gusm\~ao 
and R. Bulla for useful discussions in the early stages of this work.
It is also a pleasure to acknowledge M. Foglio for useful comments. 
This work has been partially supported by the Brazilian agency 
Con\-se\-lho Na\-ci\-o\-nal de De\-sen\-vol\-vi\-men\-to 
Ci\-en\-t\'\i\-fi\-co e Tec\-no\-l\'o\-gi\-co (CNPq), 
Max-Planck-Institut f\"ur Physik Komplexer Systeme (MPIPKS), and
Funda\c c\~ao de Amparo \`a Pesquisa do Estado de S\~ao Paulo (FAPESP). 
Part of IPT study was performed in the CRAY Y-MP2E of the 
Centro Nacional de Supercomputa\c c\~ao da Universidade Federal do 
Rio Grande do Sul, at Porto Alegre, Brazil.


\begin{references}
\bibitem{D} $D(z)\equiv\int_{-\infty}^{+\infty}
d\epsilon\rho_0(\epsilon)/(z-\epsilon)$ is the Hilbert transform of the 
uncorrelated density of states of the conduction band.
\end{references}
\end{document}